\documentclass[preprint,superscriptaddress,showpacs,aps,pre,floatfix,longbibliography]
{revtex4-1}
\usepackage[utf8]{inputenc} % allow utf-8 input
\usepackage[T1]{fontenc}    % use 8-bit T1 fonts
\usepackage{booktabs}       % professional-quality tables
\usepackage{amsfonts}       % blackboard math symbols
\usepackage{nicefrac}       % compact symbols for 1/2, etc.
\usepackage{microtype}      % microtypography
\usepackage{amssymb,amsthm}
\usepackage{amsmath,bm}
\usepackage{graphicx,epstopdf}
\usepackage{multirow}
\usepackage{subfigure}
\usepackage{setspace}
\usepackage{wrapfig}
\usepackage{comment}
\usepackage{listings}
\usepackage{amssymb,dsfont}
\usepackage{leftidx}
\usepackage{bigdelim}
\usepackage{mathrsfs}
\usepackage{blkarray,adjustbox}
\usepackage{lineno}
\usepackage{enumitem}
\usepackage{mathpazo}
\usepackage{rotating}
\usepackage{soul}
\usepackage[dvipsnames]{xcolor}
\usepackage{xr-hyper}
\usepackage{xspace}
\usepackage[pdftex, pdftitle={Article},colorlinks, pdfauthor={Author}]{hyperref} % For hyperlinks in the PDF
\usepackage[nameinlink,capitalise]{cleveref}
\hypersetup{colorlinks={true},linkcolor={darkgray},citecolor=ForestGreen}

\newcommand{\eg}{e.g.\xspace}
\definecolor{reddish}{HTML}{FBB4AE}
\definecolor{blueish}{HTML}{B3CDE3}
\definecolor{magentish}{HTML}{FF00AA}
\definecolor{greenish}{HTML}{a1d99b}

\begin{document}
\title{Dynamic predictability and spatio-temporal contexts in human mobility}

 \author{Bibandhan Poudyal}
      \affiliation{Department of Physics \& Astronomy, University of Rochester, Rochester, NY, USA}
\author{Diogo Pacheco}
    \affiliation{BioComplex Laboratory, Department of Computer Science, University of Exeter, UK}
 \author{Marcos Oliveira}
 \affiliation{BioComplex Laboratory, Department of Computer Science, University of Exeter, UK}
 \affiliation{GESIS - Leibniz Institute for the Social Sciences, Cologne, Germany}
\author{Zexun Chen}
\affiliation{University of Edinburgh Business School, UK}
\author{Hugo S. Barbosa}
	\affiliation{BioComplex Laboratory, Department of Computer Science, University of Exeter, UK}
    \affiliation{Department of Computer Science, University of Rochester, Rochester, NY, USA}	
\author{Ronaldo Menezes}
    \email[Correspondence email address: ]{r.menezes@exeter.ac.uk}% Your name
    \affiliation{BioComplex Laboratory, Department of Computer Science, University of Exeter, UK}
    \author{Gourab Ghoshal}
    \email[Correspondence email address: ]{gghoshal@pas.rochester.edu}% Your name
    \affiliation{Department of Physics \& Astronomy, University of Rochester, Rochester, NY, USA}

 %\keywords{human mobility $|$ society rhythms $|$ predictability $|$ activity signatures} 

\begin{abstract}
Human travelling behaviours are markedly regular, to a large extent, predictable, and mostly driven by biological necessities (\eg sleeping, eating) and social constructs (\eg school schedules, synchronisation of labour). Not surprisingly, such predictability is influenced by an array of factors ranging in scale from individual (\eg preference, choices) and social (\eg household, groups) all the way to global scale (\eg mobility restrictions in a pandemic). In this work, we explore how spatio-temporal patterns in individual-level mobility, which we refer to as \emph{predictability states}, carry a large degree of information regarding the nature of the regularities in mobility. Our findings indicate the existence of contextual and activity signatures in predictability states, pointing towards the potential for more sophisticated, data-driven approaches to short-term, higher-order mobility predictions beyond frequentist/probabilistic methods.
\end{abstract}

\maketitle

The understanding of mechanism governing human travelling behaviours is crucial to a variety of domains such as epidemic modelling~\cite{Balcan2009, Panos_2020, Hazarie_2021, Ansari_2022, Panos_2022},  transportation~\cite{Wang2012b, Louf2014, Lee2017, Kirkley2018}, national security~\cite{Laxhammar2014a}, urban planning~\cite{Roth2011, Zhong2014, Batty2013, Pan2013, Barthelemy2016, Bassolas2019, Mimar_2022} and a host of other applications~\cite{Barbosa2017}. Human mobility trajectories have been shown to exhibit statistical regularities at multiple scales~\cite{Gonzalez2008a, Hazarie_2020}, despite the inherent complexity that exists in the available choices for the routes of their daily travels~\cite{Aldashev_2012, Lima:2016ima}. These regularities are rooted in an array of social, spatial and temporal mechanisms, leading to visitation patterns being highly regular~\cite{Ghoshal_2014_social}. Indeed, it has been shown that a perfect algorithm can predict, with between 70-90\% certainty, an individual's future location given their prior location visits~\cite{Song2010b}, depending upon the spatio-temporal granularity of observations~\cite{Ikanovic2017}.

Moreover, human beings tend to be routine-oriented. For instance, lack of regularity in daily mobility  is linked to high levels of stress \cite{evans2002morning,quelhas2015aversive}. This change-averse behaviour leads people to favouring well-defined routines, which, in combination with stationarity~\cite{Ikanovic2017}, makes mobility trajectories quite regular~\cite{Chen_2022}. Several factors such as work schedules and physiological processes, influence mobility-related decisions; for instance, daily necessities such as sleeping and eating influence activity schedules ~\cite{Stupfel1990,Scheer2007,Toole2015,Schneider2013,Hasan2012a, Barbosa_2021}. Conversely, disruptions to such routines can completely alter the mobility trajectories (as in the recent COVID-19 lock-down measures~\cite{Santana:Second2020}), which can significantly alter mobility trajectories and therefore the associated levels of predictability~\cite{Santana2023}. 

Missing from extant measures of predictability are spatio-temporal constraints and the social embedding behind mobility regularities~\cite{DeDomenico2013}. For instance, guessing that a person will be at home on a Tuesday at 4am will most likely be a correct prediction. However,  it may be much harder to know a person's whereabouts during times in which they might not be bound by typical daily rhythms, for instance on weekends~\cite{zhang2018spatiotemporal}. Additionally, predictability is computed from an asymptotic approximation of entropy based on a non-parametric estimator~\cite{Lempel1976,Kontoyiannis1998} that  does not account for the periodic and rhythmic nature of travelling behaviours~\cite{jiang2016timegeo,cornacchia2021mechanistic}. Finally, the metric does not provide insight into the generative mechanisms that underpin the observed regularities in mobility~\cite{Barbosa2017}.

To overcome these limitations, we conduct studies on three Location-Based Social-Networks  (LBSNs) that contain the sequences of location trajectories, and propose a new perspective to mobility predictability that accounts for these missing factors, particularly connecting the observed regularities in mobility patterns to human circadian/semi-circadian routines and the types of location they visit. Furthermore, we study the temporal variations of predictability, unveiling structural patterns in their frequency and time components, which we refer to as {\em predictability states}. Our results suggest that in addition to the daily routines, regularities in mobility predictability is also marked by periods of approximately 12h and 6h, which could correspond to the second and fourth harmonics of internal circadian rhythm. These findings could suggest that factors governing  mobility-related decisions are also influenced by internal biological cycles beyond sleeping and feeding needs. This is corroborated by the predominance of 12h periods over other cycles governed by sleep/work/study routines, such as the 8h and 4h cycles. Additionally we show the role of spatial context, demonstrating heterogeneities in mobility profiles based on the types of locations that people visit. The importance of this spatial context is demonstrated through agreement between the true distribution---calculated from the full set of visitation trajectories---and a simple linear model that takes only the frequency of visiting a location-type as input. Taken together our results indicate that uncertainty and predictability in mobility patterns should be considered as a transient {\em state} of the individuals that is heavily influenced by spatio-temporal context. 
\begin{table}[t!]
    \centering
    \caption{\label{table1_datasets} Data from  location-based social networking sites.}
    \begin{tabular}{rrrc} 
    \toprule
    \multicolumn{1}{c}{Dataset} & Users & Records & Period  \\ 
    \midrule
    Brightkite~\cite{Brightkite} & $58,228$ & $4,491,143$ & Apr/2008--Oct/2010  \\ 
    Gowalla~\cite{Gowalla} & $107,002$ & $6,405,492$ & Feb/2009--Oct/2010  \\ 
    Weeplaces~\cite{Weeplaces} & $15,799$ & $7,658,368$ & Nov/2003--Jun/2011  \\ 
    \bottomrule
    \end{tabular}
    \label{tab:tab1}
\end{table}

\section*{Results}
\label{sec:results}

\subsection*{Data}
We use data from three different location-based social networking services. Brightkite and Gowalla were two popular social networking sites that existed from 2007 until 2012. Weeplaces was a website in that users could upload their check-in activities from other social network services (e.g., Facebook Places, Foursquare). These datasets contain users' check-in activity including user identification, location coordinates (i.e., latitude and longitude), and the time-stamp of the logged activity. Additionally, the Weeplaces dataset contains a description (i.e., the category) of the locations (e.g., nightlife, outdoors). The details of the data are listed in Table~\ref{tab:tab1}. For each individual in the dataset, we convert their check-in activity to trajectories described as a time-series of the form $$X = \left\{x\left(1\right), x\left(2\right), \hdots, x\left(T\right)\right\},$$ where $x(t) \in \mathcal{V}$ is a location, and $\mathcal{V}$ is the set of all visited locations by that individual.  For more details on the data, see Section S1 and Table S1. 

\subsection*{Time-independent uncertainty and predictability} 

\begin{figure}[t!]
\centering
\includegraphics[width=\textwidth]{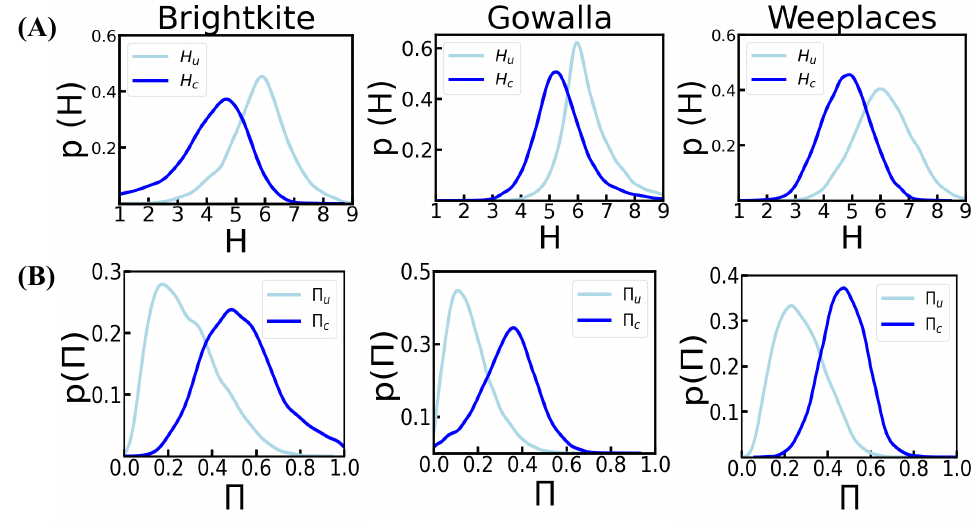}
% Legend (350 words max).
\caption{
{\bf Time-independent entropy and predictability} {\bf A} The entropy accounting for only visitation frequencies $H_u$ and incorporating the temporal sequence of visitations $H_c$. {\bf B} The corresponding predictability values calculated via Eq.~\eqref{eq:predict}. In all cases, accounting for the sequence reduces uncertainty from 6 to 5 bits and increases predictability from $20\%$ to $40\%$.
}
\label{fig:max_predictability}
\end{figure}

We begin our analysis by examining the information contained in the location trajectories of all individuals in each of the datasets. When accounting for only frequency of location visitations, the degree of uncertainty in capturing the future locations of a trajectory given past observations is encoded in the Shannon entropy (measured in bits)
\begin{equation}
H_u = -\sum_{i\in \mathcal{V}} p(x) \log_2 p(x),
\label{eq:Shannon}
\end{equation}
where $p(x)$ is the probability to visit location $x$. The subscript refers to the fact that this is the uncorrelated entropy, given that no information on the sequence of location visits is considered. Accounting for both the visitation frequency and the temporal sequence of location visits, we use a non-parametric estimator~\cite{Lempel1976,Kontoyiannis1998} termed the entropy rate, given by the expression
\begin{equation}
    H_c= \frac{N \log_2 N}{\sum\limits_{i=1}^N\Lambda_{i}},
    \label{eq:lempel}
\end{equation}
where $N$ is the number of locations visited by an individual and 
$\Lambda_{i}$ is the length of the shortest trajectory sub-sequence beginning at position $i$ not seen previously. In the absence of any structure in the sequence Eq.~\eqref{eq:lempel} reduces to Eq.~\eqref{eq:Shannon}. In Fig.~\ref{fig:max_predictability}{\bf A} we plot the results for the datasets finding that in all cases $H_u \approx 6$ bits and $H_c \approx 5$ bits. This indicates that accounting for the temporal sequence reduces the possible space of future location visits from $2^6$ to $2^5$ possible locations indicating the presence of temporal correlations in visits between locations. 

The entropy rate can be converted to a measure of predictability $\Pi$  using Fano's inequality~\cite{Cover2006} to define the upper bound of how often an ideal predictive algorithm can correctly guess the next location visit, given prior history. 
This is calculated by inverting 
\begin{equation}
H_{u,c} \leq B(\Pi_{u,c}) + (1 - \Pi_{u,c})\log_{2}(n-1),
\label{eq:predict}
\end{equation}
where $n$ is the number of \emph{distinct locations visited} and $B(x)$ is the binary entropy function capturing the entropy of a simple Bernoulli trial (in this case achieving maximal predictability or not). The metric mathematically bounds the performance of all real predictive methods given an information sources inferred uncertainty.  The corresponding results are shown in Fig.~\ref{fig:max_predictability}{\bf B} and mirrors the trends seen for the entropy; including information about the sequence of location visits, over and above the frequency of visiting those locations, increases the predictability from around 20\% to 40\%. Given the inherent information in the sequence, in what is to follow, we restrict the analysis to the correlated entropies and predictability $H_c,  \Pi_c$.

\subsection*{Temporal Predictability}

\begin{figure}[t!]
\centering
\includegraphics[width=14cm]{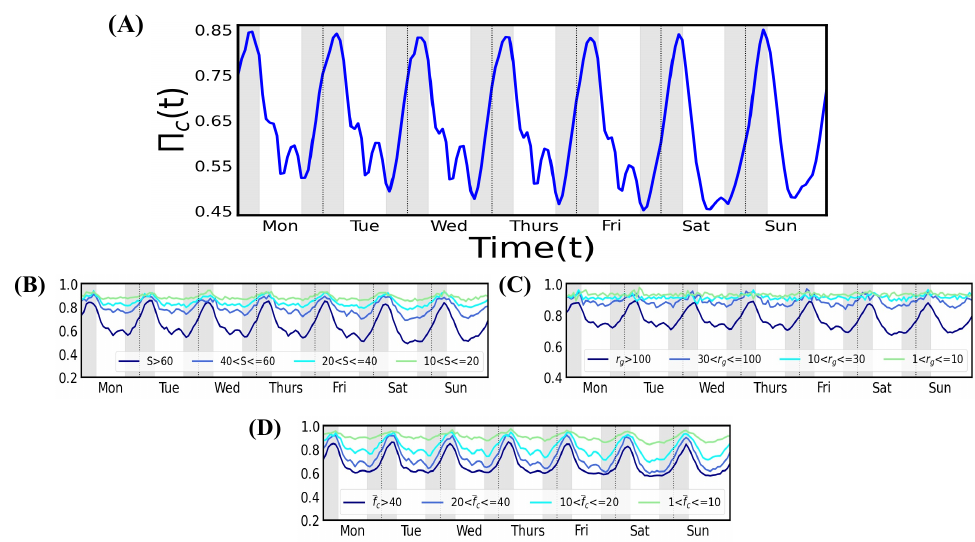}
% Legend (350 words max).
\caption{\textbf{Time-dependent predictabilities.} {\bf A} The predictability disaggregated with respect to time for Weeplaces. The trajectory of each individual is split into time slots representing the time of the week. We create $168$ slots (i.e., $24\, \text{hours} \times 7$ days of the week) and define $X_{t=t_0}$ as a random variable representing the places that an individual visits at the time slot $t=t_0 \in [1, \dots, 168]$.{\bf B} The temporal predictability disaggregated with respect to the number of unique locations visited $S$. {\bf C} Now disaggregated with respect to geographical coverage as measured by the radius of gyration $r_g$ in units of kilometer. {\bf D} Finally, disaggregated with respect to the average frequency of monthly check-ins $\bar f_c$. Across all users we see daily peaks (4-5am) and secondary peaks (12-5pm) of predictability throughout the time series. A increasing diversity in activity (more locations visited, larger areas covered, and more check-ins) the predictability progressively decreases. 
}
\label{fig:circadian}
\end{figure}

Human activity routines are characterised by temporal regularities with time and frequency components. For instance, usual working hours tend to recur every 24h (i.e., frequency) with changes during the weekends (i.e., time). Thus, it is reasonable to believe that mobility regularities in both time and frequency domains should manifest themselves in the predictability profile of an individual. Thus, in order to extract the temporal variation of the predictability, instead of considering the complete visitation sequence of individuals, we analyse individuals at different moments of the week. Specifically, we split the trajectory of each individual into time slots representing the time of the week. We create $168$ slots (i.e., $24\, \text{hours} \times 7$ days of the week) and define $X_{t=t_0}$ as a random variable representing the places that an individual visits at the time slot $t=t_0 \in [1, \dots, 168]$. 

In Fig.~\ref{fig:circadian}{\bf A} we plot the time-series of the predictability $\Pi_c (t)$ for Weeplaces---the corresponding plots for Brightkite and Gowalla are shown in Figs. S2 and S3. As the figure indicates, when disaggregating with respect to time, the predictability exhibits considerable variation both from day-to-day as well as within a given day. For instance we see a progressive decrease in predictability from Monday to Friday, which then picks up again and peaks on the weekend. Within a day there is considerable variation, with the predictability of an individual fluctuating between 50-85\%, with individuals being more predictable at night. Finally, we see a clear 24h periodicity in the predictability profile with high predictability periods during evenings and at night (peaking around 4-5am) and troughs at mid-day. This feature is seen across all three datasets.   
 
 We note that as in other human facets, human mobility and social interactions tend to exhibit a high degree of heterogeneity. If on the one hand most people use their mobile phones a few times a day, some users, on the other hand, make hundreds of calls a day. Some individuals are more active in the sense of travelling more frequently, to a diversity of locations, as well as traveling longer distances. One would expect this spatio-temporal context to influence the predictability profiles in different ways. To test for this effect we disaggregate the data in terms of three different metrics of activity: number of unique location visited $S$; geographical coverage measured by the \textit{radius of gyration} $r_g$~\cite{Gonzalez2008a}; and monthly recurrent frequency by averaging the number of check-ins per month $\bar f_c$. Figure S1 shows the distribution of these quantities for all the datasets indicating a right-skewed heavy-tailed distribution for all three measures in line with previous observations~\cite{Gonzalez2008a, Song2010b}.
 
The influence of the observed heterogeneity in the activity metrics on the predictability is plotted in Figs.~\ref{fig:circadian}{\bf B-D}. First, we note that the temporal trends are maintained even while disaggregating with respect to the activity metrics. However the range in predictability varies with respect to the levels of activity. For instance, those who visit between 10-20 unique locations have an effectively flat temporal profile across the days of the week, as well as within a day ($\Pi_c \approx 90\%$). In contrast those that visit greater than 60 unique locations show much more temporal variability with lower levels of predictability ($ 50\% \leq \Pi_c \leq 80\%$). Additionally one sees a sharp drop in predictability when comparing populations visiting less than 60 locations to those visiting more than this number. A similar trend is seen for geographical coverage, where once again a flat trend is exhibited by populations venturing not more than 10kms from their residence, whereas those travelling more than 100 kms, experience the same variability as those visiting more than 60 unique locations. Now one sees a sharp drop in $\Pi_c$ when comparing populations travelling more than 100kms to those that travel less. Finally, the trend is also mirrored when measuring the frequency of check-ins $\bar f_c$, the difference being the absence of any sharp drops in $\Pi_c$ with a more smooth decrease between the segmented populations. Figures S2 and S3 indicate the same trends for BrightKite and Gowalla. Across all datasets the predictability decreases with more diversity in activity (as measured by $S$, $r_g$, and $\bar f_c$). That is, it is harder to predict the mobility of users with more \emph{diverse} routes irrespective of whether the diversity is measured in terms of the number of unique locations visited, the geographical area explored, or the amount of monthly data (traces). Taken together the results indicate an unexplored facet of uncertainty and predictability; stating that ``an individual is $80\%$ predictable''  % in terms of their behavior then
must be interpreted in an \emph{averaged} sense. Missing from this is the instantaneous changes in a person's \emph{predictability state} over time.

\begin{figure}[t!]
\centering
\includegraphics[width=\textwidth]{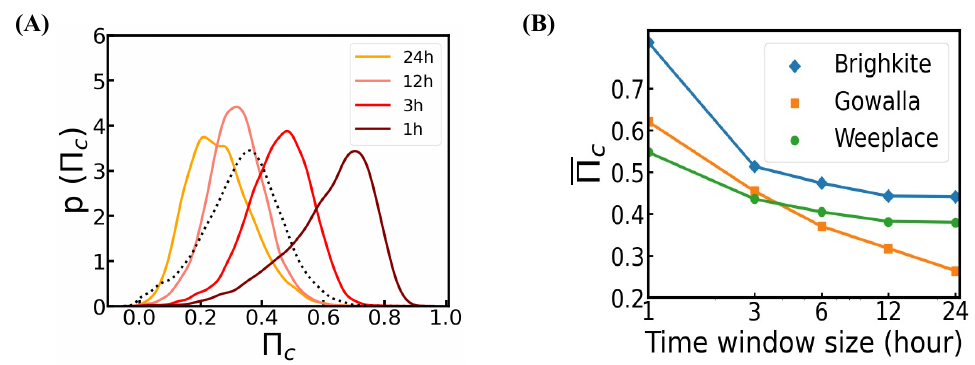}
% Legend (350 words max).
\caption{\textbf{The effect of temporal windows of observation}
{\bf A} The distribution of $\Pi_c$ as a function of window-size for Weeplaces. The dashed curve represent the baseline distribution of $\Pi_c$ when taking into account the full trajectory of individuals. {\bf B} The modes of the distributions plotted as a function of window-size for all three datasets. Horizontal dashed lines indicate the saturated value of $\Pi_c$.}
\label{fig:time_granularity}
\end{figure}

Indeed, one would also expect this to vary with respect to the observation window of an individual's trajectory. For instance, if we observe an user only for hour, they are likely to visit only a few locations. Given the lack of diversity in routes, the predictability should be high (as indicated by Fig.~\ref{fig:circadian}). Conversely the longer the observation window, the more the number of locations visited and therefore the predictability necessarily should decrease. It is interesting to consider whether there is a saturation in this decrease in $\Pi_c$. In 
Fig.~\ref{fig:time_granularity}{\bf A } we plot the distribution of $\Pi_c$ as a function of the temporal window of observations for Weeplaces. As a baseline, the curve for the predictability considering the full set of trajectories (absent any window) is shown as a dashed line. As the temporal window decreases, from a week-day (seven 24-hour bins per week) to a week-day-hour (hundred and sixty-eight 1-hour bins), the average predictability increases as expected. In Fig.~\ref{fig:time_granularity}{\bf B} we plot mean of the distributions as a function of the window-size finding a saturation in the curve at 24h for all three datasets. Thus it appears that an observation window of 24h is a reasonable approximation to considering the full trajectory in terms of uncovering the distribution of $\Pi_c$.

\begin{figure}[t!]
\centering
\includegraphics[width=\textwidth]{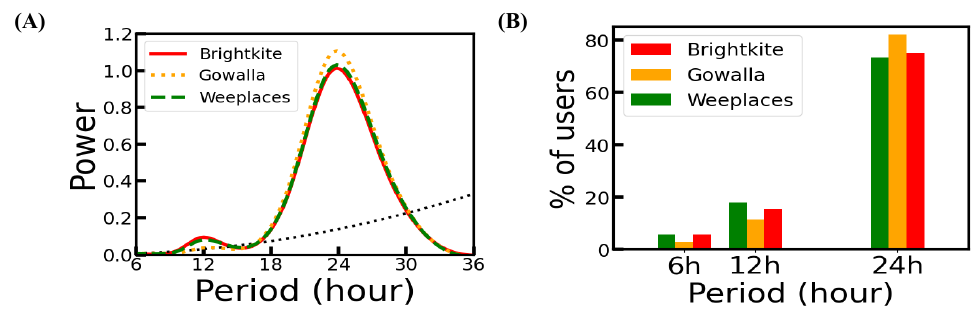}
% Legend (350 words max).
\caption{\textbf{Temporal modes of predictability} {\bf A}. Estimated global wavelet power spectrum showing peaks at 24h (circadian) and 12h (circasemedian) as well as a non-significant peak at 6h. The dashed black line represents the statistical significance of the mode as measured by Eq. S3.
\textbf{B.} Stacked bar chart for the most strong component period (24, 12, 6 hours) of each individual .}
\label{fig:wavelet}
\end{figure}

\subsection*{Temporal and frequency modes in predictability}
The temporal variation of $\Pi_c$ in Fig.~\ref{fig:circadian} suggests the possible presence of additional frequency modes in addition to the clear 24h periodicity. To uncover this one can use the continuous wavelet transform to describe the regularities in the time series of the individuals~\cite{Percival1995, Torrence1998}.  Wavelet analysis reveals the frequency components of signals just like the Fourier transform, but it also identifies where a certain frequency exists in the temporal or spatial domain (See Section S4 for details of the method). 
Fig.~\ref{fig:wavelet}{\bf A} shows the wavelet power spectrum for each of the datasets (the dashed line indicates the statistical significance Eq. S3). Not surprisingly, the figure reveals the circadian period (approximately 24h) as the most prominent component of the predictability regularity. Additionally, and somewhat surprisingly, the 12h component is the second-strongest component (i.e., the circasemidian period). Given working schedules and sleeping cycles, one might have expected a mode around the 8h period, however, the third-strongest component is centred approximately around the 6h regime during the day, even though the signal is not statistically significant at the population level. We note that the power spectrum is essentially identical across all three datasets, indicating the robustness of these modes in determining the observed temporal variation in predictability. 
Conducting the wavelet analysis on individual-level data, one can represent the percentage of user in each dataset that have either the 24h,12h or 6h mode as their strongest components. The results are shown in Fig.~\ref{fig:wavelet}{\bf B} indicating that across all datasets around 75\% of users have the 24h mode as their strongest component. Around 20\% of users have instead the 12h mode as their strongest component whereas a negligible fraction have the 6h mode.

\subsection*{Spatial context of predictability}
Thus far we have investigated the temporal context of uncertainty in mobility behaviour. It stands to reason, that there is also a spatial-- or location--based context that influences movement patterns. 
For example, it is plausible that people are more predictable about their workplace or residence, as compared to locations that represent leisure and entertainment activities such as visiting restaurants and museums. This is a function of those types of locations being less frequently visited as compared to those that are driven by the daily work schedules. 

If one were to use the types of locations as a proxy for spatial context, then one can scan the sequence of trajectory visits and group each location into categories. The Weeplaces dataset contains a standardised set of eight location tags: \textit{Home/Work, Education, Entertainment, Food, Travel, Shops, Outdoors and Nightlife}. Restricting the analysis of trajectories to each location-type, one can compare it to the baseline distribution of $\Pi_c$ when considering all types of locations. For instance, a context-dependent trajectory for food could be: \textit{breakfast-place-A}, \textit{lunch-place-B}, \textit{coffee-place-C}, etc; while a full baseline trajectory could be: \textit{home-place-X}, \textit{breakfast-place-A}, \textit{work-place-Y}, \textit{lunch-place-B}, \textit{coffee-place-C}, \textit{gym-place-Z}, etc. 

\begin{figure}[t!]
\centering
\includegraphics[width=\textwidth]{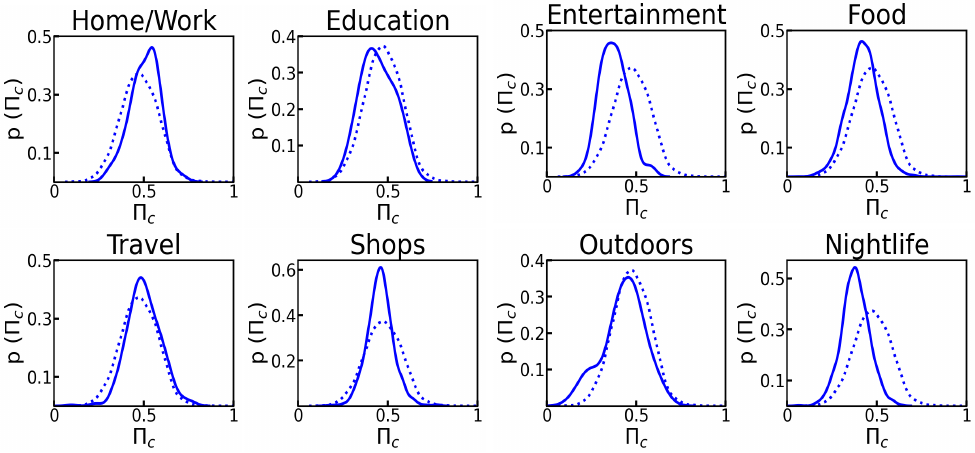}
% Legend (350 words max).
\caption{\textbf{Spatial context of predictability} In all subplots, the dotted-lines represent the distribution of $\Pi_c$ given their full trajectories for Weeplaces. Solid-lines represent distributions for the same individuals but when limiting their trajectories to locations of distinct types.
}
\label{fig:activity_type}
\end{figure}

The results are shown in 
Fig.~\ref{fig:activity_type} with the baseline distribution shown as a dashed line. The figure clearly indicates the role of a spatial-context in the distribution of $\Pi_c$. For instance, as expected,  activities related to \textit{Home/Work, Travel, Shops} are more predictable than the baseline; \textit{Education, Outdoors} essentially mirror the baseline, whereas activities corresponding to \textit{Food, Entertainment, Nightlife} have peaks markedly below the baseline.  

The observed difference in predictability as a function of location-context behooves us to investigate the extent to which the frequency of which particular location types are visited, has a bearing on the overall distribution of $\Pi_c$. Note this is different from merely considering the frequency distribution of visiting a \emph{particular location} which is the primary input to Eq.~\eqref{eq:Shannon}. Instead, here we coarse-grain these locations into categories and investigate whether this feature can be a good estimator for $P(\Pi_c)$. To that effect, we consider a linear regression model by using the relative frequency that an individual stayed at places from each category as an input. For instance, we investigate whether knowing that a person goes shopping often, informs us about this person's overall predictability. We train the model with the calculated predictability as an independent variable and the relative frequencies of the categories as the dependent variable. 

\begin{figure}[t!]
\centering
\includegraphics[width=\textwidth]{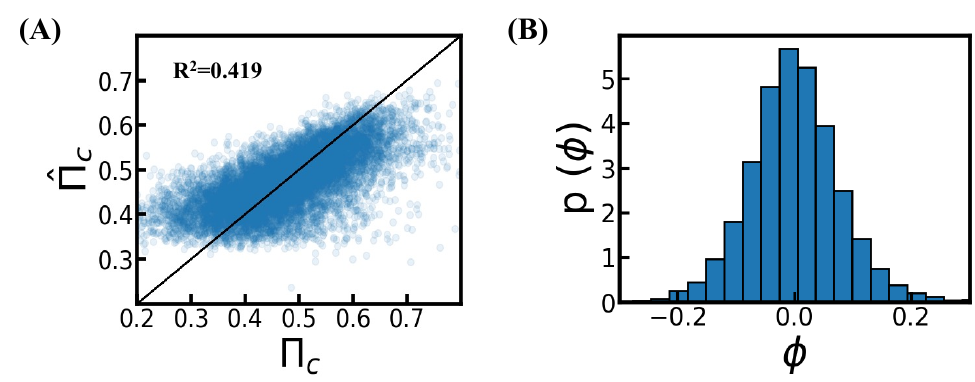}
% Legend (350 words max).
\caption{\textbf{Estimating predictability from location-context}. {\bf A} The estimated predictability $\hat \Pi_c$ using a linear model with frequency of visiting location-types as input, plotted against the true distribution of $\Pi_c$ in the Weeplace dataset ($R^2=0.419$). {\bf B} The residuals are normally distributed and centred around zero.}
\label{fig:estimating_predictability}
\end{figure}

Using this linear model, we estimate the predictability denoted by $P(\hat{\Pi}_c)$  (calculated from coarse-grained trajectories, e.g., 80\% home/work, 18\% food, 1\% nightlife, etc) in Fig.~\ref{fig:estimating_predictability}{\bf A} and plot it against the \emph{true} distribution $P(\Pi_c)$ (calculated from the full trajectories over multiple days, e.g., home, restaurant, work, restaurant, etc.). As the figure indicates, this simple model does a reasonable job of estimating the true distribution ($R^2=0.419$) with the residuals being  
normally distributed and centred around zero (Fig.~\ref{fig:estimating_predictability}{\bf B}). (The coefficients of the model are shown in Tab. S2).
The results are instructive in a couple of ways. First, the role of spatial context is clearly important given that the linear model is a reasonable approximation to the true distribution; sxecond, it also  implies that coarse-grained information about individuals is enough to capture the qualitative trends in mobility uncertainty, without recourse to the fine-grained full set of mobility trajectories. 

\section*{Discussion}
\label{sec:discussion}

Meeting a citizenry's needs require governments, industries, and other stakeholders to be able to plan for demands (e.g., hospital admissions, public transportation, store opening times). The predictability of human movement is at the heart of planning, hence more accurate modelling should inform better decision-making in terms of public policy. Extant research, models uncertainty and predictability in mobility trajectories as as aggregated value for individual considering their full set of mobility trajectories as an input. This neglects the spatio-temporal factors influencing mobility, such as the temporal variations for different times of the day and days of the week, as well as in space. As our results indicate, the predictability is greatly influenced by the window of observation, the length of the observation, the diversity of activities undertaken by an individual, as well as the location of the individual. Consequently, a more accurate way of modelling predictability is to consider it as a \emph{transient state}, while taking into account the spatio-temporal context. 

The role of spatial context has some practical considerations. As shown, coarse-graining the locations into types and measuring the frequency with each location type visited is a reasonable proxy for the predictability that is extracted from the mobility trajectories, that requires finely-grained data with enough sampling.  Indeed, typically high-resolution mobility data is hard to come by and is restricted to only a few regions in the world. Our results show that much can be learnt even with low-resolution information. On the other hand, there is increasing access to higher resolution data, such as the ones being collected as part of ``track-and-trace'' systems in certain countries. In such instances, our framework of combining mobility traces with spatio-temporal context can lead to more accurate characterisations. Indeed, many governments and companies (as part of ``data-for-good'' efforts) are starting to open their datasets to scientists which will naturally lead to better urban analytics, including human dynamics modelling.

The main implications of this work is that planing of activities related to human mobility (e.g. city events, epidemic modelling, road maintenance) need to consider time-space variations of individual activities. Furthermore, during periods of restrictions, such as the COVID-19 pandemic, the understanding and characterisation of these time-spatial variations can aid governments to make the correct decisions. In 2020 and 2021, many governments imposed curfew/lockdown measures to citizens after certain hours (e.g., Spain, Colombia) in a ``blanket'' way. Effective curfews depend on the time and the location, and the consideration of such variations can lead to a better approach where not all areas are treated equally. Using predictability as a state and using the states for planning could lead to more just/equitable outcomes.

Our work of course has limitations; given the analysis was conducted on a particular type of dataset (LBSN's) that vary in their spatiotemporal resolution, geographical coverage, and sampling rates. It will be instructive to conduct this analysis in different regions of the world and using other sources of mobility information such as censuses, GPS data and Call Data Records. Nevertheless, given that we conducted the analysis on three different datasets, finding similar trends points towards the results holding up to scrutiny in different settings.

\section*{Acknowledgement}

We thank Brooke Foucault-Welles for useful discussions on the paper.

% Bibliography
\bibliographystyle{naturemag}
\bibliography{manuscript}

\clearpage

\end{document}

\makeatletter
\renewcommand{\maketitle}{\bgroup\setlength{\parindent}{0pt}
	\begin{flushleft}
{\huge Supplementary material}\\[0.3cm]
		{\large\textbf \@title}
		
		\@author
		
	\end{flushleft}\egroup
}
\makeatother
\title{Dynamic predictability and spatio-temporal contexts in human mobility}

\newcommand{\floor}[1]{\lfloor #1 \rfloor}

\author{Bibandhan \surname{Poudyal}, Diogo \surname{Pacheco}, Marcos \surname{Oliveira}, Zexun \surname{Chen}, Hugo S. \surname{Barbosa}, Ronaldo \surname{Menezes} and Gourab \surname{Ghoshal}}                         

\date{}

\maketitle

\tableofcontents

\newpage

\clearpage

\section{Dataset pre-processing}
FIG. \ref{Data_preprocessing} below shows the distribution of three different quantitites capturing activity patterns in mobility trajectories. The distribution of the radius of gyration $r_g$ resembles a truncated power-law distribution and is qualitatively identical among all three datasets . The truncation for Weeplaces occur around $10^3$km whereas for Brightkite and Gowalla, it occurs around $10^4$km. The distribution of no of distinct locations $S$ and the average monthly check-ins frequency $\bar f_c$ shows few characteristic differences between the dataset. First, Brighkite users have lesser probability to check-in new locations and their average monthly check-in frequency is also likely to be lesser than in the other datasets. Second, Weeplaces exhibits a heavier tail  both in no of distinct locations and average monthly check-in frequency indicating that Weeplaces user are more active than other users.
\begin{figure}[h!]
\centering
    \includegraphics[width=\textwidth]{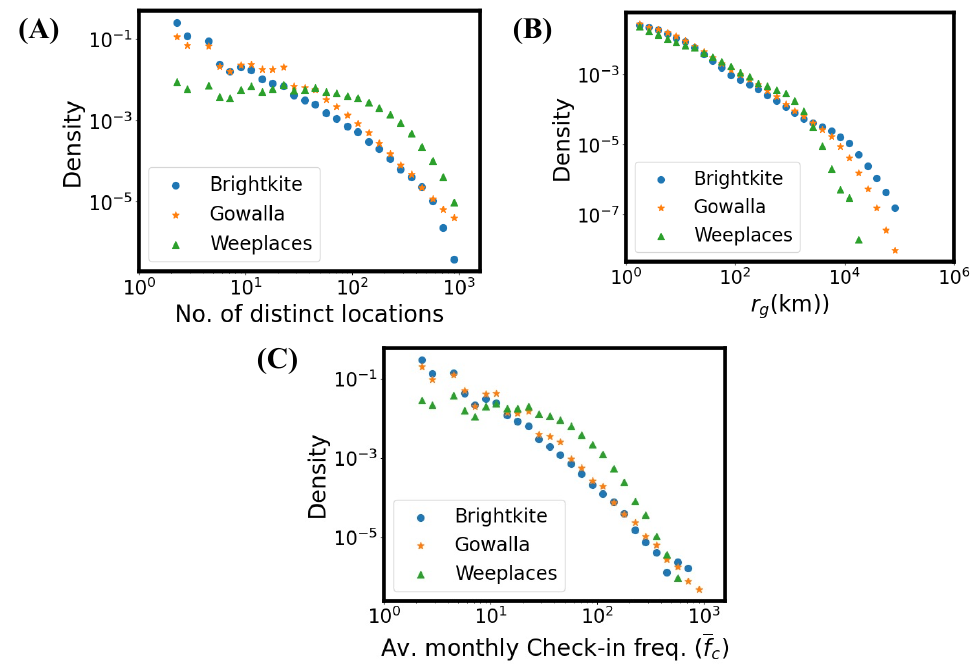}
    \caption{\textbf{Distribution of selected mobility quantities} \textbf{(a)} Distribution of number of distinct locations visited by all users in the dataset. \textbf{(b)} Distribution of radius of gyration of all users in the dataset. \textbf{(c)} Distribution of average monthly check-in frequency of all users in the dataset.}
    \label{Data_preprocessing}
\end{figure}

\begin{figure}[h!]
\centering
    \includegraphics[width=\textwidth]{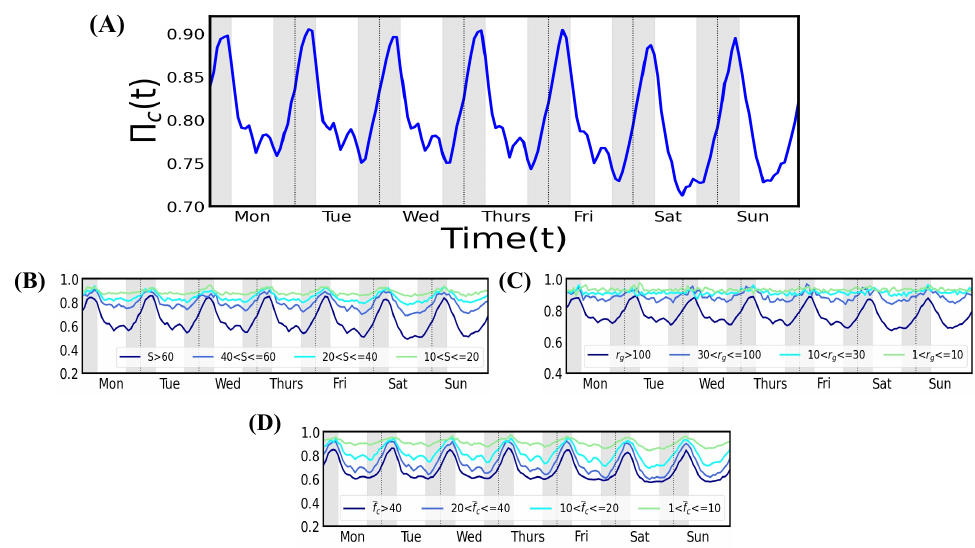}
    \caption{\textbf{(a)} The average predictability $\Pi_c$ across all Brightkite users exhibit daily peaks (4-5am) and secondary peaks (12-5pm) of predictability throughout the time series. For all datasets, $\Pi_c$ is lower for groups with \textbf{(b)} more unique locations $S$, \textbf{(c)} more radius of gyration $r_g$ and \textbf{(d)} more average monthly check-ins $\Bar{f}_c$.}
    \label{Brightkite}
\end{figure}

\begin{figure}[h!]
\centering
    \includegraphics[width=\textwidth]{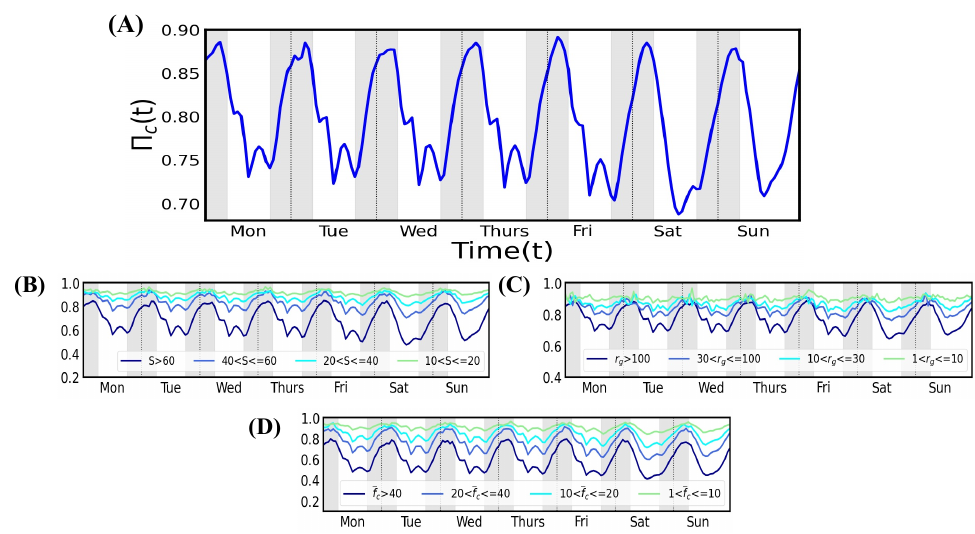}
    \caption{\textbf{(a)} The average predictability $\Pi_c$ across all Gowalla users exhibit daily peaks (4-5am) and secondary peaks (12-5pm) of predictability throughout the time series. For all datasets, $\Pi_c$ is lower for groups with \textbf{(b)} more unique locations $S$, \textbf{(c)} more radius of gyration $r_g$ and \textbf{(d)} more average monthly check-ins $\Bar{f}_c$.}
    \label{Gowalla}
\end{figure}

Given the sparsity and differences in the dataset, the pre-processing of dataset was done with the objective to set a threshold for our data such that the predictability estimator saturates and yield a robust output. During the longitudinal analysis of the predictability along different days of the week, we find that the predictabilty for no. of distinct location (S)>60 resembles most accurately the overall predictability as show in FIG. 2, FIG. \ref{Brightkite} and FIG. \ref{Gowalla}. Thus, for all other analysis in the manuscript, with the exception of Fig. 2, we use $S>60$ as a threshold to compute the results.

\begin{table}[h!]
\centering
        \caption{Summary of the preprocessed dataset.}
\scalebox{0.8}{
\begin{tabular}{ | p{2cm}| p{6cm}| p{2cm}| p{4cm}|}
 \hline
\textbf{Dataset} & \textbf{Total Check-ins/Call Records} & \textbf{Users} & \textbf{Distinct Placeid}\\
 \hline
BrightKite & 2,279,147 & 4,401 & 667,549 
\\
\hline
Gowalla & 4,061,522 & 15,874 & 2,454,544 
 \\
 \hline
Weeplaces & 5,589,296 & 10,388 & 1,947,392
 \\
 \hline
\end{tabular}}
\label{tab-indicators}
\end{table}
\clearpage

\section{Brightkite predictability}
    
\clearpage

\section{Gowalla predictability}
    
\clearpage

\section{Wavelet analysis}
To understand the temporal dimension of human mobility predictability, we use the continuous wavelet transform to describe the regularities in the time series of the individuals. With the wavelet transform, we extract both  time and frequency components from a time series. The method has a long history of successful applications to a variety of domains such as climate prediction~\cite{Nalley2016}, digital image processing~\cite{antoine1993image}, and crime dynamics analysis~\cite{oliveira2018spatio}. 
The wavelet transform decomposes a time series using functions, called wavelets that dilate (scale) to capture different frequencies and that translate (shift) in time to include changes \textit{with} time. 
We can define the wavelet transform of a discrete sequence $Y = \{y(1), y(2), \ldots,y(N)\}$ having observations with a uniform step $\delta t$ as the following: 
\begin{equation}
	{{\rm W}_{Y}(s,n)=\sqrt{\dfrac{\delta t}{s}}\sum_{t=1}^N y(t)\psi^{*}\bigg[\frac{(t-n)\delta t}{s}\bigg]},
\end{equation}
where the `$*$' denotes the complex conjugate and $s$ is the wavelet scale. 
The wavelet transform can be seen as the cross-correlation between the time series $y\left(t\right)$ and a set of functions $\psi^{*}_{s,\tau}\left(t\right)$, distributed over time and having different widths~\cite{Cazelles2007}.
By varying the scale $s$ and translating over time (i.e., varying $n$), we have a representation of the amplitude of the different periodic features of $Y$ and how they vary with time.

%  The transform ${\rm W}_{Y^c}\left(s,n\right)$ describes the contribution of a periodicity (scale) $s$ at different moments $n$ in the time series $X$. 
 
 To examine the overall periodicity in $Y$, we evaluate the average of ${\rm W}_{Y^c}\left(s,n\right)$ over $n$:
\begin{equation}
{\overline{{\rm W}}^2\left(s\right) = \frac{1}{N}\sum_{n=1}^{N}|{\rm W}_{Y^c}\left(s,n\right)|^2},
\label{eq:temporal_averaged_power}
\end{equation}
called the global wavelet spectrum, which provides us with an unbiased estimate of the true power spectrum~\cite{Percival1995}. We analyse its statistical significance by using the method developed by Torrence and Compo~\cite{Torrence1998}, which tests the wavelet power against a null model that generates a background power spectrum $P_k$. The test is given by: 
\begin{align}
	{D\left(\frac{|\overline{W}_X\left(s, n\right)|^2}
    {\sigma_X^2} < p\right) = \frac{1}{2}P_k\chi^2_\nu,}
	\label{TorrenceCompoTestPowerSpectrum}
\end{align}
where $\nu=2$ for complex wavelets (our case)~\cite{Torrence1998}.

%\section{Effects of time-granularity in the limits of uncorrelated predictability}
%    
%\begin{figure}[h!]
%\centering
%    \includegraphics[width=16cm]{figures/Bib_Figure_S3.pdf}
%    \caption{(A) The effect of time window on the distribution of uncorrelated predictability where dashed line represents the overall distribution. (B) The effect of time window on the average uncorrelated predictability.}
%\end{figure}

\clearpage

\clearpage

\section{Linear regression model}

% \begin{table}[h!]
%     \centering
%     \caption{OLS}
% \begin{tabular}{lclc}
% \toprule
% \textbf{Dep. Variable:}    &     pi\_max      & \textbf{  R-squared:         } &     0.419   \\
% \textbf{Model:}            &       OLS        & \textbf{  Adj. R-squared:    } &     0.419   \\
% \textbf{Method:}           &  Least Squares   & \textbf{  F-statistic:       } &     1071.   \\
% \textbf{Date:}             & Thu, 07 Sep 2023 & \textbf{  Prob (F-statistic):} &     0.00    \\
% \textbf{Time:}             &     19:33:59     & \textbf{  Log-Likelihood:    } &    11694.   \\
% \textbf{No. Observations:} &        10388      & \textbf{  AIC:               } & -2.337e+04  \\
% \textbf{Df Residuals:}     &        10380      & \textbf{  BIC:               } & -2.331e+04  \\
% \textbf{Df Model:}         &           7      & \textbf{                     } &             \\
% \bottomrule
% \end{tabular}
% \label{stab:OLS}
% \end{table}

\begin{table}[h!]
    \centering
        \caption{Regression Results Coefficients}
\begin{tabular}{lrcrcrr}
\toprule
                       & \textbf{coef} & \textbf{std err} & \textbf{t-statistic} & \textbf{95\% CI}  \\
\midrule
\textbf{constant}      &       0.4216    &        0.002     &    218.979      &           [0.418, 0.425 ]    \\
\textbf{Education}     &       0.2264  &        0.010     &    22.159  &       [0.206, 0.246 ]    \\
\textbf{Entertainment} &      -0.2233  &        0.014     &   -15.430  &    [-0.252, -0.195]     \\
\textbf{Food}          &      -0.0861  &        0.005     &    -15.754  &  [-0.097, -0.075]     \\
\textbf{Home/Work}     &       0.3234  &        0.005     &    65.292  &  [0.314, 0.333]     \\
\textbf{Nightlife}     &      -0.0938  &        0.008     &    -12.378  &  [-0.109, -0.079 ]    \\
\textbf{Outdoors}      &       0.0379  &        0.013     &     2.877  & [0.0012, 0.064]    \\
\textbf{Shops}         &       0.1007  &        0.007     &    13.963  & [0.087, 0.115]   \\
\textbf{Travel}        &       0.1365  &        0.006     &    22.337  &  [0.124, 0.148]     \\
\bottomrule
\end{tabular}
\label{stab:OLS2}
\end{table}

% \begin{table}[h!]
%     \centering \caption{OLS Regression Results Statistics}
% \begin{tabular}{lrlr}
% \toprule
% \textbf{Omnibus:}       & 411.18 & \textbf{  Durbin-Watson:     } &    1.928  \\
% \textbf{Prob(Omnibus):} &  0.000 & \textbf{  Jarque-Bera (JB):  } &   912.292  \\
% \textbf{Skew:}          &  0.251 & \textbf{  Prob(JB):          } & 7.91e-199  \\
% \textbf{Kurtosis:}      &  4.362 & \textbf{  Cond. No.          } & 1.79e+16  \\
% \bottomrule
% \end{tabular}
% \label{stab:OLS3}
% \end{table}

% % Notes: \newline
% %  [1] Standard Errors assume that the covariance matrix of the errors is correctly specified. \newline
% %  [2] The smallest eigenvalue is 6.62e-29. This might indicate that there are \newline
% %  strong multicollinearity problems or that the design matrix is singular.

\bibliographystyle{naturemag}
\bibliography{manuscript}